# Decreasing log data of multi-tier services for effective request tracing

Bo Sang[1,2], Jianfeng Zhan[1], Guanhua Tian[1,2]
[1]Institute of Computing Technology, Chinese Academy of Sciences
[2]The Graduate University of Chinese Academy of Sciences
sangbo@ncic.ac.cn

## Abstract

*Previous work shows request tracing systems help understand and debug the performance problems of multi-tier services. However, for large-scale data centers, more than hundreds of thousands of service instances provide online service at the same time. Previous work such as white-box or black box tracing systems will produce large amount of log data, which would be correlated into large quantities of causal paths for performance debugging. In this paper, we propose an innovative algorithm to eliminate valueless logs of multi-tiers services. Our experiment shows our method filters 84% valueless causal paths and is promising to be used in large-scale data centers.*

## 1. Introduction

Request tracing systems [2] [3] help understand and debug performance problems of large concurrent services, since they gain insight into component activities and correlate activities into *causal paths*. A *causal path* is a sequence of interaction activities of components of multi-tier services with causal relationship caused by an individual request.

Frequent requests cause the tracing systems to produce huge amount of logs of *component activities* and correlate them into large quantity of causal paths. For example, using the tracing method of Microsoft labs, a simple e-commercial system [2] generates 150K events per minute, which amounts to 10 M log data per minute. Usually, a data center has 10 thousands or even more nodes being deployed with multi-tier services. If we use this tracing method to debug the performance problems of multi-tier services on this scale for *one minute*, the tracing system needs to analyze at least *0.1 TB log data*, which results in unacceptable cost.

A typical web service system has three major tiers to deal with requests: a web server tier, an active web pages server tier and a database tier. Using the tracing method, we obtain two kinds of causal paths: *simple causal paths* capture only interactions between clients and the first tier; *complex causal paths* reflect interactions among clients and at least two tiers. Since simple causal paths are a part of complex causal paths and complex causal paths include more interaction information than simple causal paths, the complex ones are more valuable than the simple ones for performance debugging of multi-tier services. If we can distinguish valuable complex causal paths from valueless simple causal paths, it will improve the effectiveness of request tracing.

Our work improves our previous work PreciseTracer [3]. We develop a new algorithm to distinguish complex request paths from simple ones and eliminate those valueless logs. We also improve the tool of constructing precise paths [3] and decrease the quantity of causal paths to only 20%.

## 2. Background

In our previous work [3], we collect four types of activities: activities identifying the start (*BEGIN*) and end (*END*) of servicing a request, activities of sending (*SEND*) or receiving (*RECEIVE*) a message on components through kernel instrumentation. For each activity, we log four attributes: activity type, timestamp, context identifier (*hostname, program name, ID of process, ID of thread*) and message identifier *(IP of sender, port of sender, IP of receiver, port of receiver)*. Then we transmit all logs of activities to the correlation server and correlate them into causal paths.

Statistic data [2] indicates that 80% requests only require static content on web servers. In experiments, we also discover a considerable percent of paths (more than 600 in 1000) only pass through the first tier. We only need the rest complex causal paths for debugging and should eliminate the simple ones.

## 3. Solution
### 3.1. Elimination algorithm

On each node, PreciseTracer transforms the original format of activities into more understandable n-ary tuples (*tuple records*) to describe activities. This stage is called as *Transformation*. As shown in Fig.1, we make elimination at *Transformation* stage and eliminate the log data of interaction activities between clients and the first tier, which are correlated into simple causal paths. In Fig.1 the first tier is Web server, so we eliminate the valueless logs on the node running Web server.

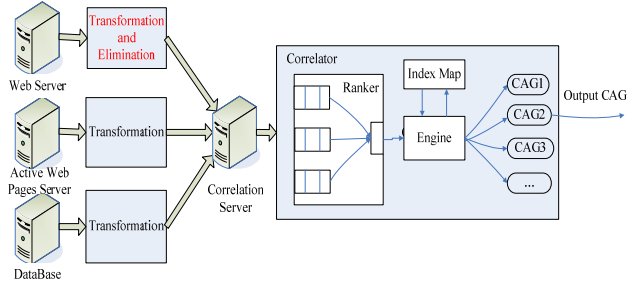

Fig.1. The architecture of PreciseTracer.

In most of multi-tier service systems, threads of components deal with only one request at a time. Thus if we can discover requests producing simple causal paths at the beginning, we can eliminate all logs of activities related to their Thread IDs. So the first message from the client is the only one we can utilize. We discover that the size of the first message from the client in simple causal paths is less than that in complex ones. Using *k-means clustering method* [1], we compute a threshold value of the size of the first message to distinguish simple causal paths from complex ones. We use a *map<Tid, state>* to record a thread's current state *(start, simple, complex, end)*. Our algorithm deals with all original logs by classifying them into four cases:

Case 1: The log records BEGIN activity. We mark the related Tid's mapping *state = start* and transform it to tuple record;

Case 2: The log records RECEIVE activity and its receiving port is 80 and its related Tid's mapping *state = start*. If the size of the message is over the threshold value, we mark its related Tid's mapping *state = complex* and transform it to tuple record, otherwise *state = simple*;

Case 3: The log records END activity. We mark its related Tid's mapping *state = end* and transform the log;

Case 4: The log records other kinds of activities. If its related Tid's mapping *state = complex*, we transform it into tuple record, otherwise pass through it without any operations.

There are three advantages of our algorithm: (1) It eliminates large amount of valueless logs; (2) It eliminates logs on local server node and reduce the requirement for bandwidth; (3) It decrease the amount of final causal paths.

## 4. Evaluation
## 4.1. Experiment and setup

We make experiments on RUBiS [3]. Web server (Apache), active web pages server (JBoss) and database (MySQL) are deployed on three nodes separately. All nodes are SMP with two PIII processors (1G) and run RedHat 4.1.1-30 Linux with SystemTap as instrumentation tool to gather logs.

The experiment is divided into three steps: 1. Sampling instrumentation, which lasts for 1.5 minutes; 2. Transforming original logs and eliminating logs of simple causal paths; 3. Correlating causal paths.

## 4.2. Results

|   | Original logs | Tuple Records | Causal Paths |
|---|---|---|---|
| No elimination | 9.5M | 11M | 12373 |
| Elimination | 9.5M | 2.5M | 1997 |

**Table 1**

After elimination, tuple records reduce by 77% and final causal paths reduce by 84%. Almost all final causal paths after elimination are complex causal paths.

## 5. Conclusion

The large amount of log data in data centers is the main challenge for the application of request tracing. Our work concludes features of multi-tier service systems and develops an algorithm to decrease causal paths significantly by eliminating valueless log data. Our further work will focus on developing a more universal method for complex systems.

## 6. Acknowledgements

This paper is supported by the NSFC (Grant No. 60703020).